\documentclass[apj]{emulateapj}
\usepackage{apjfonts}

\newcommand{\hii}         {\mbox{\rm \ion{H}{2}}}

\newcommand{\kms}         {km~s$^{-1}$}

\newcommand{\ha}          {\mbox{H$\alpha$}}

\def\spose#1{\hbox to 0pt{#1\hss}}
\def\lta{\mathrel{\spose{\lower 3pt\hbox{$\mathchar"218$}}
     \raise 2.0pt\hbox{$\mathchar"13C$}}}
\def\gta{\mathrel{\spose{\lower 3pt\hbox{$\mathchar"218$}}
     \raise 2.0pt\hbox{$\mathchar"13E$}}}

\shorttitle{Constraints on CSM Around SN 2007af}
\shortauthors{Simon et al.}
\slugcomment{Accepted for publication in ApJL}

\begin{document}

\title{Constraints on Circumstellar Material Around the Type Ia
  Supernova 2007af\altaffilmark{1,2}}

\author{Joshua D. Simon\altaffilmark{3}, Avishay
  Gal-Yam\altaffilmark{3,4}, Bryan E. Penprase\altaffilmark{5}, Weidong
  Li\altaffilmark{6}, Robert M. Quimby\altaffilmark{7}, Jeffrey
  M. Silverman\altaffilmark{6}, Carlos Allende Prieto\altaffilmark{7},
  J. Craig Wheeler\altaffilmark{7}, Alexei V. Filippenko\altaffilmark{6},
  Irene T. Martinez\altaffilmark{5}, Daniel J. Beeler\altaffilmark{5},
  and Ferdinando Patat\altaffilmark{8}}

\altaffiltext{1}{Some of the data presented herein were obtained at
  the W. M. Keck Observatory, which is operated as a scientific
  partnership among the California Institute of Technology, the
  University of California, and NASA.  The Observatory was made
  possible by the generous financial support of the W. M. Keck
  Foundation.}

\altaffiltext{2}{Based in part on observations obtained with the
  Hobby-Eberly Telescope, which is a joint project of the University
  of Texas at Austin, the Pennsylvania State University, Stanford
  University, Ludwig-Maximilians-Universit{\" a}t M{\" u}nchen, and
  Georg-August-Universit{\" a}t G{\" o}ttingen.}

\altaffiltext{3}{Department of Astronomy, California Institute of
  Technology, 1200 E. California Blvd., MS 105-24, Pasadena, CA 91125;
  jsimon@astro.caltech.edu, avishay@astro.caltech.edu}

\altaffiltext{4}{Astrophysics Group, Faculty of Physics,
                 Weizmann Institute of Science, 76100 Rehovot, Israel}

\altaffiltext{5}{Department of Physics and Astronomy,
                 Pomona College, 610 N. College Ave., Claremont, CA  91711;
                 penprase@dci.pomona.edu}

\altaffiltext{6}{Department of Astronomy,
                University of California, Berkeley, CA 94720-3411;
                weidong@astro.berkeley.edu, 
                jsilverman@astro.berkeley.edu, alex@astro.berkeley.edu}

\altaffiltext{7}{McDonald Observatory and Department of Astronomy, 
                University of Texas, Austin, TX  71782;
                quimby@astro.as.utexas.edu,
                callende@astro.as.utexas.edu, wheel@astro.as.utexas.edu}

\altaffiltext{8}{European Southern Observatory, Karl Schwarzschild Str. 2,
                 D-85748 Garching bei M{\"u}nchen, Germany; fpatat@eso.org}

\begin{abstract}

\citeauthor{patat} recently inferred the existence of circumstellar
material around a normal Type Ia supernova (SN) for the first time,
finding time-variable Na~I~D absorption lines in the spectrum of SN
2006X.  We present high-resolution spectroscopy of the bright SN~Ia
2007af at three epochs and search for variability in any of the Na~D
absorption components.  Over the time range from 4~d before to 24~d
after maximum light, we find that the host-galaxy Na~D lines appear to
be of interstellar rather than circumstellar origin and do not vary
down to the level of 18~m\AA\ (column density of $2 \times
10^{11}$~cm$^{-2}$).  We limit any circumstellar absorption lines to
be weaker than $\sim10$~m\AA\ ($6 \times 10^{10}$~cm$^{-2}$).  For the
case of material distributed in spherically symmetric shells of radius
$\sim10^{16}$~cm surrounding the progenitor system, we place an upper
limit on the shell mass of $\sim(3 \times 10^{-8})/X$~M$_{\odot}$, where
$X$ is the Na ionization fraction.  We also show that SN 2007af is a
photometrically and spectroscopically normal SN~Ia.  Assuming that the
variable Na~D lines in SN 2006X came from circumstellar matter, we
therefore conclude that either there is a preferred geometry for the
detection of variable absorption components in Type Ia supernovae, or
SN 2007af and SN 2006X had different types of progenitor systems.

%Ca H and K
%Halpha
\end{abstract}

%binaries: close/general/symbiotic?
%stars: evolution?
%stars: winds, outflows?
%white dwarfs?

\section{INTRODUCTION}
\label{intro}

Type Ia supernovae (SNe~Ia) are currently the only distance indicator
that can be used effectively out to cosmological distances
\citep[e.g.,][]{riess98,perlmutter99,leibundgut04,filippenko05,riess07}.
Therefore, understanding the nature of these explosions and any
potential systematics that may be present is of great importance to
cosmology.  However, we still do not know what the progenitor systems
of SNe~Ia are, and observations suggest that there may be at least two
physically different progenitor classes
\citep[e.g.,][]{mannucci05,sb05,mannucci06,sullivan06,quimby07}, as
well as some peculiar objects \citep[e.g.,][although see
  \citealp{benetti06}]{li01,li03,hamuy03}.

\citet{patat} have recently made a possible breakthrough in the study
of SN~Ia progenitors by optically detecting circumstellar material
(CSM) in a SN~Ia for the first time.  Using high-resolution spectra of
SN 2006X spanning from just before maximum light to four months later,
they showed that at least four distinct components of the Na~I~D
absorption lines varied with time until $\sim 2$ months
post-explosion.  Although similar behavior has been seen in Milky Way
stars and is generally attributed to small interstellar clouds moving
across the line of sight \citep[e.g.,][]{wf01}, in SN 2006X the lack
of time evolution in the corresponding \ion{Ca}{2} H \& K absorption
features at the same velocities probably rules out that interpretation.
Instead, \citeauthor{patat} conclude that the variable absorption is
from circumstellar clouds in the progenitor system that were ionized
by the radiation from the supernova and recombined several weeks
later; because \ion{Na}{1} has a much lower ionization potential than
\ion{Ca}{2}, the \ion{Na}{1} line profiles can change without an
accompanying effect in the \ion{Ca}{2} lines if the ionizing radiation
has an appropriate spectrum.  These results appear to indicate a
single-degenerate progenitor for SN 2006X with a red-giant companion.

Multiple-epoch high-resolution spectroscopy is available for only one
previous SN~Ia, the peculiar SN 2000cx \citep{patat07b}, so it is not
yet known whether the time evolution seen in SN 2006X is common.
Assuming that the variable absorption is related to material from the
SN progenitor, if the behavior of SN 2006X is not universal then
either there must be geometric effects that limit the visibility of
the absorption to certain lines of sight (e.g., near the orbital plane
of the progenitor system) or there are multiple progenitor systems for
SNe~Ia.

In this \emph{Letter}, we present high-resolution spectra of the Type
Ia SN 2007af obtained at $-4.3$, $+16.6$, and $+23.7$ d relative to
maximum light.  We use these data to test the \citet{patat} model,
searching for variability in the Na~D absorption features.

\section{OBSERVATIONS AND DATA REDUCTION}
\label{observations}

SN 2007af was discovered by K. Itagaki on 2007 March 1.84 \citep[UT
  dates are used throughout this paper;][]{ni07}.  A spectrum obtained
on 2007 March 4.34 showed that SN 2007af was a SN~Ia at least a week
before peak brightness \citep{salgado07}.  The host galaxy of the
supernova is NGC~5584, an Scd galaxy with a recession velocity of
1638~\kms\ \citep{koribalski04}.

\subsection{High-Resolution Spectroscopy}

We observed SN 2007af with the ARCES echelle spectrograph
\citep{wang03} on the ARC 3.5~m telescope at Apache Point Observatory
(APO) on 2007 March 10.  We obtained two 1440~s exposures covering the
spectral range 3200--10,000~\AA\ at a spectral resolution of $R
\approx 33,000$ and a signal-to-noise ratio (S/N) of 36 per pixel at
the wavelength of the redshifted Na~D lines.  The data were reduced in
IRAF\footnote{IRAF is distributed by the National Optical Astronomy
  Observatories, which are operated by the Association of Universities
  for Research in Astronomy, Inc., under cooperative agreement with
  the National Science Foundation.} with the {\sc echelle} package
using standard procedures.

We also observed SN 2007af with the high-resolution spectrograph
\citep[HRS;][]{hrs} on the Hobby-Eberly Telescope (HET) on 2007 March
31.  The spectrograph was in its $R = 30,000$ mode, with a
2\arcsec-diameter fiber and the 316 $\ell$/mm grating centered at
6948~\AA, providing nearly complete wavelength coverage from 5095 to
8860~\AA.  We obtained 4 spectra totaling 3100~s of exposure time and
reached a combined S/N of 87 per pixel.  The HRS data were reduced in
IRAF with the {\sc echelle} package.

Finally, we observed SN 2007af with the HIRES spectrograph
\citep{vogt94} on the Keck I telescope on 2007 April 7, over the range
3150--6000~\AA.  We obtained a single 900~s exposure through a
$7\farcs0 \times 0\farcs861$ slit, yielding $R \approx 48,000$ and S/N
= 47 per pixel.  The HIRES data were processed with the MAKEE data
reduction package.

\subsection{Imaging and Low-Resolution Spectroscopy}

SN 2007af was the target of extensive photometric follow-up
observations with the 0.76~m Katzman Automated Imaging Telescope
\citep[KAIT;][]{li00,filippenko01}, continuing for over 4
months since its discovery.  No pre-explosion $BVRI$ images are available 
for subtraction of the host-galaxy light, but the supernova occurred
relatively far out in the disk, away from significant contaminating
features.  We used the {\sc daophot} package \citep{stetson87} in IRAF
to perform point-spread function photometry of SN 2007af relative 
to various field stars in the KAIT images, which were calibrated on 
five photometric nights with KAIT and the Nickel 1~m telescope at Lick 
Observatory.

We also obtained low-resolution spectra of SN 2007af with the Kast
spectrograph (Miller \& Stone 1993) on the Shane 3~m telescope at Lick
on 2007 March 13, April 10, and June 14, which were reduced in IRAF
following normal procedures.
%4173.045
%4200.896
%4265.814

\section{RESULTS}
\label{results}

\subsection{Light Curve and Low-Resolution Spectra}

\begin{figure}[t!]
\epsscale{1.24}
\plotone{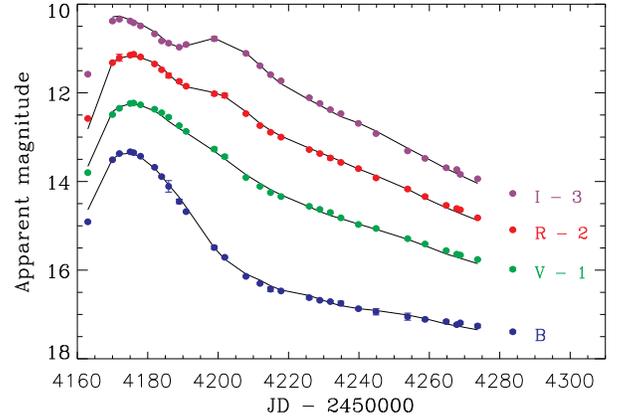}
\caption{$BVRI$ light curves of SN 2007af, from KAIT data.
  Photometric uncertainties are indicated by the plotted error bars,
  which in most cases are smaller than the displayed data points.  The
MLCS fits are shown by the black curves.}
\label{lightcurves}
\end{figure}

We display $BVRI$ light curves of SN 2007af in Figure
\ref{lightcurves}.  We fitted the photometric data with the latest
version of the multicolor light-curve shape method
\citep[MLCS2k2;][]{jrk07} to determine the parameters of the
supernova.  We find that the time of $B$-band maximum was 2007 March
14.76 (JD = 2,454,174.26), with an uncertainty of 0.12~d.  The derived
line-of-sight extinction to the SN is $A_{V} = 0.39 \pm 0.06$~mag,
with an extinction law of $R_{V} = 2.98 \pm 0.33$ (the Milky Way
foreground reddening is 0.039~mag; \citealt{sfd98}).  The distance
modulus to SN 2007af is $(32.06 - 5\log{\mbox{H}_{0}/72}) \pm
0.06$~mag, giving the SN an absolute magnitude of $M_{V} = (-19.28 +
5\log{\mbox{H}_{0}/72}) \pm 0.08$.  The luminosity/light-curve-shape
parameter is $\Delta = -0.04 \pm 0.02$, and the MLCS2k2 reduced
$\chi^{2}$ value of 0.53 indicates an excellent fit.  Photometrically,
SN 2007af appears normal in every respect.

\begin{figure}[t!]
\epsscale{1.24}
\plotone{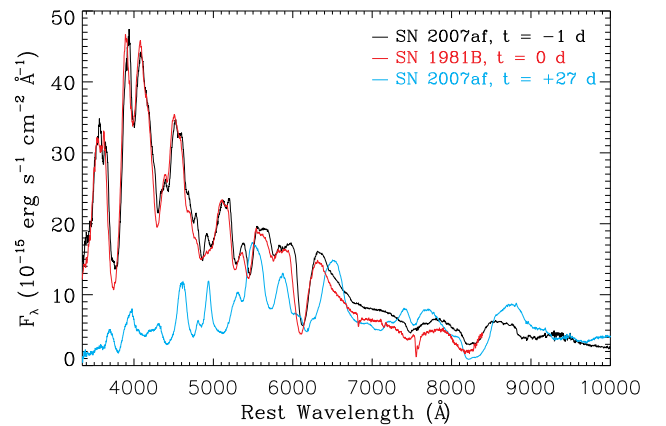}
\caption{Low-resolution spectrum of SN 2007af one day before maximum
  light (black curve), compared to a scaled spectrum of the
  prototypical SN~Ia 1981B at maximum light (red curve; data from
  \citealt{branch83}; telluric absorption is present at $\sim 6830$~\AA\
  and $\sim 7560$~\AA).  The cyan curve shows the spectrum of SN 2007af 
  four weeks later, for comparison.}
\label{spectra}
\end{figure}

In Figure \ref{spectra} we show the spectrum of SN 2007af one day
before maximum light, along with a template spectrum of SN 1981B at a
similar epoch.  Analysis with the Superfit spectral fitting code
\citep{howell05} indicates that SN 2007af is a very typical SN~Ia,
comparable to archetypal events such as SNe 1981B and 1989B
\citep[e.g.,][]{filippenko97}.

\subsection{Sodium D Absorption Lines at High Resolution}
\label{na}

\begin{figure}[t!]
\epsscale{1.2}
\plotone{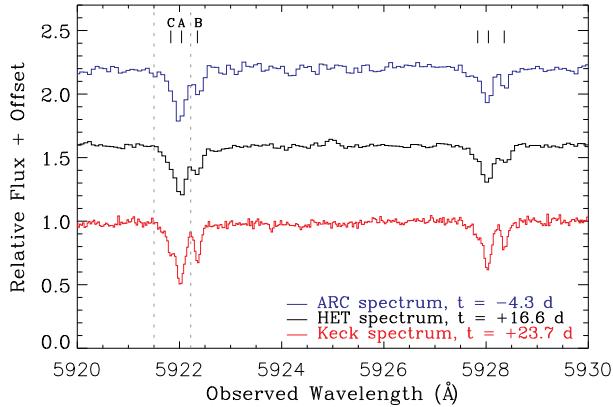}
\caption{High-resolution spectra of host-galaxy Na~D absorption lines
  in SN 2007af.  The HET and ARC spectra are offset by 0.6 and 1.2,
  respectively, in the vertical direction for clarity.  The tickmarks
  labeled ``A,'' ``B,'' and ``C'' at the top of the plot indicate the
  wavelengths of the three absorption components, and the dotted gray
  lines show the wavelength range used to measure the EW for
  components A and C.  Note that the Keck spectrum has significantly
  higher spectral resolution than the other spectra (see \S
  \ref{observations}).}
\label{highres}
\end{figure}

We display the high-resolution spectra of SN 2007af around the
host-galaxy Na~D lines in Figure \ref{highres}.  At least two
absorption components are present.  In the ARC and HET spectra, the
shape of the blue wing of the absorption profile (as well as the
residuals from two-component fits) suggest that a third component at
slightly lower velocity might be blended with the stronger absorption
line, and examination of the higher-resolution Keck spectrum confirms
this.

We fit the Na~D$_{1}$ and D$_{2}$ absorption lines in each spectrum
with three Gaussian components using a Levenberg-Marquardt
least-squares fit algorithm.\footnote{We obtain equivalent results
  with Lorentzian or Voigt profiles.}  Each absorption component is
given its own depth, width, and wavelength, but we hold the wavelength
separation between the D$_{1}$ and D$_{2}$ lines fixed at its known
value and we force the D$_{1}$/D$_{2}$ depth ratio to be the same for
each component.  Including the continuum level, there are therefore 11
fitting parameters for 123--308 spectral channels (depending on the
dispersion of each spectrum).  We list the fit results in Table
\ref{na_fits}.

While the equivalent widths (EWs) of components A and C appear to
change somewhat across the three spectra, in fact these components are
so badly blended in the ARC and HET spectra that the individual EWs
are not very well determined.  Instead of measuring the EWs of these
components separately with the Gaussian profile fits, we directly
integrate the spectra from 5921.50~\AA\ to 5922.22~\AA.  The combined
EWs of components A and C in the three epochs are $139\pm6$~m\AA,
$137\pm3$~m\AA, and $148\pm3$~m\AA.\footnote{All \ion{Na}{1} EWs in
  this paper are given for the D$_{2}$ component.  The D$_{1}$/D$_{2}$
  depth ratio (and therefore the D$_{1}$/D$_{2}$ EW ratio) is 0.64.}
We therefore conclude that there is no change in these absorption
lines to a 3$\sigma$ limit of 18~m\AA\ over the course of the
observations.  We estimate that the \ion{Na}{1} column density for
components A and C is $9 \times 10^{11}$~cm$^{-2}$ (assuming a Doppler
parameter of $b \approx 8$~\kms; see Table \ref{na_fits}), and the
change in the column between 4~d before and 24~d after maximum light
is no larger than $2 \times 10^{11}$~cm$^{-2}$.
The strength of component B, which is reasonably well separated from
the other two, does not change within the uncertainties over the
course of the observations, and we estimate a \ion{Na}{1} column
density of $(3 \pm 0.2) \times 10^{11}$~cm$^{-2}$ for this component.
We place 5$\sigma$ upper limits on the presence of additional
unresolved absorption components at other velocities of 15~m\AA,
8~m\AA, and 7~m\AA\ (9, 5, and $4 \times 10^{10}$~cm$^{-2}$) for the
ARC, HET, and Keck spectra, respectively.  In contrast, the
time-variable features in SN 2006X reached \ion{Na}{1} column
densities of up to $10^{12}$~cm$^{-2}$ \citep{patat}.

\subsection{Calcium H\&K Absorption Lines at High Resolution}
\label{ca}

The APO and Keck spectra extend far enough to the blue that we also
detect \ion{Ca}{2} H\&K absorption lines in both the Milky Way and
NGC~5584.  We fit the four sets of lines simultaneously using the same
technique as in \S \ref{na}.  As with the Na lines, the host-galaxy
absorption can be well fitted with two or possibly more components.
Their velocities and relative strengths, however, differ significantly
from those of the Na absorption components.  The Ca absorption systems
have heliocentric velocities of 1643 and 1613~\kms, and the
higher-velocity system is approximately 3 times as strong as the
lower-velocity system (EWs of $160\pm4$ and $49\pm3$~m\AA\ in the
\ion{Ca}{2} K line, respectively).  These results suggest that the
various absorbing clouds along this line of sight have different
abundances and/or ionization states.  Although the higher-velocity
component appears somewhat deeper and the lower-velocity component
appears narrower in the Keck spectrum than in the APO data, we do not
detect any statistically significant changes in the absorption line
equivalent widths between the two epochs.

\subsection{Limits on \ha\ Emission}
\label{ha}

The ARC and HET spectra cover the expected wavelength of the
redshifted \ha\ line.  We detect an \ha\ emission line in both spectra
with a velocity of 1641~\kms\ and a full width at half-maximum (FWHM)
of 16~\kms.  In the earlier, but lower S/N, ARC spectrum ($t =
-4.3$~d) we measure an EW for the \ha\ emission of $45\pm11$~m\AA, and
in the HET spectrum ($t=16.6$~d) we measure an EW of $87\pm4$~m\AA.
At these epochs, the supernova had $R$-band magnitudes of $13.32 \pm
0.02$ and $13.85 \pm 0.02$, respectively.  Within the uncertainties,
therefore, the \ha\ flux did not change between the two observations.
While \ha\ emission could be a signature of CSM, we also find weak
[\ion{N}{2}]~$\lambda$6583 and [\ion{S}{2}]~$\lambda$6717 features at
the same velocity.  The presence of these additional lines, the lack
of variability in the \ha\ flux, and the close agreement between the
host-galaxy velocity and the \ha\ velocity suggest instead that the
emission is coming from a nearby \hii\ region.

Assuming that the detected emission lines come from the host galaxy
rather than the supernova, we place an upper limit on the
\ha\ emission from the SN of 22~m\AA\ (5$\sigma$ limit for a line
width of 50~\kms) at $t=16.6$~d and 33~m\AA\ at $t=-4.3$~d.  Although
the high-resolution spectra are not flux-calibrated, we can estimate
the \ha\ limits in flux units by scaling the continuum flux level to
the observed $R$-band magnitudes.  We calculate an upper limit to the
\ha\ flux of $\sim2 \times 10^{-16}$~erg~cm$^{-2}$~s$^{-1}$
($L_{\ha} = 2 \times 10^{37}$~erg~s$^{-1}$).  Using the model
of \citet{cumming96}, the corresponding upper limit on the mass-loss
rate in the progenitor system is not very restrictive
($\lesssim10^{-4.5}$~M$_{\odot}$~yr$^{-1}$).  Several other nearby SNe~Ia
have comparable or better limits from \ha\ observations
\citep{cumming96,mattila05,patat}.

\section{DISCUSSION AND CONCLUSIONS}
\label{discussion}

Our observations of Na absorption lines in \S \ref{na} can be used to
estimate the amount of hydrogen along the line of sight to SN 2007af.
The total measured \ion{Na}{1} column density is $1.2 \times
10^{12}$~cm$^{-2}$.  The corresponding hydrogen column is $(8.1 \times
10^{17})/X$~cm$^{-2}$ for a solar Na abundance of 12 + log(Na/H)$ =
6.17$ \citep{ags05}, where $X = $~N(\ion{Na}{1})/N(Na) is the Na
ionization fraction.  If this material were all located in the SN
progenitor system in a thin shell with a radius of $\sim10^{16}$~cm
(as suggested by \citealt{patat} for SN 2006X), the shell mass would
be $(8.5 \times 10^{-7})/X$~M$_{\odot}$.

However, because of the lack of variations in the column density of
the detected absorption lines and their close agreement with the 
host-galaxy velocity, the absorbing gas is more likely associated with 
the interstellar medium of NGC~5584.  In that case, the relevant
calculation for the mass of the CSM is based on the upper limits for
additional absorption components.  Using the 5$\sigma$ upper limits on
the \ion{Na}{1} column density of 9, 5, and $4 \times
10^{10}$~cm$^{-2}$ for the three spectroscopic epochs and the
assumptions given above, we find corresponding upper limits on the
shell masses of $6.4/X$, $3.6/X$, and $2.9/X \times
10^{-8}$~M$_{\odot}$, respectively.  For comparison, \citet{patat}
estimated a shell mass of $7.1 \times 10^{-7}$~M$_{\odot}$ for SN
2006X with the same model.  If the circumstellar Na ions have not yet
fully recombined by the time of our observations, then the true CSM
mass could be significantly higher, but for SN 2006X substantial
recombination had occurred within 2 weeks of maximum light.

There are two primary ways to interpret the absence of variable
absorption features in SN 2007af.  First, the progenitor system may
differ from the red-giant companion in a recurrent nova model that
\citet{patat} proposed for SN 2006X.  In particular, \citet{wang07}
found that SN 2006X has the highest expansion velocity ever measured
for a SN~Ia, which raises the possibility that varying \ion{Na}{1} D
absorption features (and hence the proposed recurrent nova
progenitors) could be associated with the subgroup of high velocity
gradient SNe~Ia \citep{benetti05}.  If this conjecture is correct, it
would suggest that the companion to the SN 2007af progenitor was
either a main-sequence star, a subgiant, or another C/O white dwarf,
leaving the progenitor system relatively free of circumstellar gas and
dust.

Alternatively, the observational differences between the Na~D lines in
SNe 2007af and 2006X could be a result of the system geometry.  If the
mass lost by recurrent novae is concentrated in the orbital plane of
the system, as appears to be the case for RS Oph
\citep{obrien06,bode07}, then lines of sight that do not pass near the
orbital plane would miss most of the circumstellar material.  Such a
configuration would allow the progenitor systems of SNe 2006X and
2007af to be the same, as long as the viewing geometry is different.

Monitoring of the Na~D lines in a larger sample of SNe~Ia will clarify
the geometrical constraints, including possible additional geometries
for the CSM.  It will also allow us to examine whether the presence or
absence of variable absorption is related to any other observed
properties of the objects.

\acknowledgements{This publication was based in part on observations
  obtained with the APO 3.5-m telescope, which is owned and operated
  by the Astrophysical Research Consortium.  The HET is named in honor
  of its principal benefactors, William P. Hobby and Robert E. Eberly.
  The authors also wish to acknowledge the very significant cultural
  role and reverence that the summit of Mauna Kea has always had
  within the indigenous Hawaiian community.  We are most fortunate to
  have the opportunity to conduct observations from this mountain.  We
  thank the referee, Nikolai Chugai, for constructive comments.
  J.D.S. acknowledges the support of a Millikan Fellowship provided by
  Caltech.  R.Q. and J.C.W.  are supported in part by NSF grant
  AST--0707769, and C.A.P. acknowledges support from NASA under grants
  NAG5--13057 and NAG5--13147. A.V.F.'s group at U.C. Berkeley is
  supported by NSF grant AST--0607485, the Sylvia and Jim Katzman
  Foundation, and the TABASGO Foundation.  We thank Wal Sargent and
  Michael Rauch profusely for the Keck spectrum, HET resident
  astronomer Heinz Edelmann for carrying out the HET observations, and
  George Becker for helpful conversations.}

\begin{deluxetable}{lcccccccccc}
%\tablenum{1}
%\rotate
\tablewidth{0pt}
\tabletypesize{\scriptsize}
\tablecolumns{10}
\tablecaption{Na D Fit Results}
\tablehead{
  \multicolumn{1}{c}{} & \multicolumn{3}{c}{Component A}  & 
  \multicolumn{3}{c}{Component B} & \multicolumn{3}{c}{Component C} &
  \multicolumn{1}{c}{} \\
  \multicolumn{1}{c}{} & 
  \multicolumn{3}{c}{$\overline{\phm{SpanningSpanningSpanningspan}}$} &
  \multicolumn{3}{c}{$\overline{\phm{SpanningSpanningSpanningspan}}$} & 
  \multicolumn{3}{c}{$\overline{\phm{SpanningSpanningSpanningspan}}$}  &
  \multicolumn{1}{c}{} \\
\colhead{Spectrum} & \colhead{Velocity\tablenotemark{a}} & 
\colhead{FWHM} & \colhead{EW} & 
\colhead{Velocity\tablenotemark{a}} & \colhead{FWHM} & 
\colhead{EW} & \colhead{Velocity\tablenotemark{a}} & 
\colhead{FWHM} & \colhead{EW} 
& \colhead{$\tilde{\chi}^{2}$} \\
\colhead{} & \colhead{[\kms]} & 
\colhead{[\kms]} & \colhead{[m\AA]} & 
\colhead{[\kms]} & \colhead{[\kms]} & 
\colhead{[m\AA]} & \colhead{[\kms]} & 
\colhead{[\kms]} & \colhead{[m\AA]} & \colhead{} }
\startdata 
ARC  & $1632.2 \pm 0.5$ & $11.6 \pm 2.1$ & $80 \pm 25$ & $1650.4 \pm 0.6$ & $9.5$\tablenotemark{b} & $43 \pm 8$  & $1625.0 \pm 7.6$ & $26.9 \pm 7.5$ & $62 \pm 37$  & 0.74 \\
HET  & $1633.2 \pm 0.7$ & $11.3 \pm 1.2$ & $87 \pm 18$ & $1649.1 \pm 0.4$ & $12.3 \pm 0.9$ & $54 \pm 4$  & $1622.5 \pm 3.4$ & $15.3 \pm 4.1$ & $45 \pm 16$  & 1.68 \\
Keck & $1632.6 \pm 0.2$ & $10.1 \pm 0.4$ & $104 \pm 4$ & $1649.4 \pm 0.1$ & $7.3 \pm 0.3$ & $47 \pm 3$  & $1621.7 \pm 0.4$ & $8.5 \pm 0.8$ & $34 \pm 4$  & 1.33 \\
\enddata
\tablenotetext{a}{Velocities are in the heliocentric frame.}
\tablenotetext{b}{According to the fit this line is unresolved, so we
  cannot establish an uncertainty on the line width.}
\label{na_fits}
\end{deluxetable}


\begin{thebibliography}{}

\bibitem[Asplund et al.(2005)]{ags05} Asplund, M., Grevesse, N., \&
  Sauval, A.~J.\ 2005, in ASP Conf. Ser. 336, Cosmic Abundances as
  Records of Stellar Evolution and Nucleosynthesis,
  ed. T.~G.~Barnes~III \& F.~N.~Bash (San Francisco, ASP), 25

\bibitem[Benetti et al.(2006)]{benetti06} Benetti, S., Cappellaro, E.,
  Turatto, M., Taubenberger, S., Harutyunyan, A., \& Valenti,
  S.\ 2006, \apjl, 653, L129

\bibitem[Benetti et al.(2005)]{benetti05} Benetti, S., et al.\ 2005,
  \apj, 623, 1011

\bibitem[Bode et al.(2007)]{bode07} Bode, M.~F., Harman, D.~J.,
  O'Brien, T.~J., Bond, H.~E., Starrfield, S., Darnley, M.~J., Evans,
  A., \& Eyres, S.~P.~S.\ 2007, ApJ, in press (preprint at ArXiv
  e-prints, 706, arXiv:0706.2745)

\bibitem[Branch et al.(1983)]{branch83} Branch, D., Lacy, C.~H.,
  McCall, M.~L., Sutherland, P.~G., Uomoto, A., Wheeler, J.~C., \&
  Wills, B.~J.\ 1983, \apj, 270, 123

\bibitem[Cumming et al.(1996)]{cumming96} Cumming, R.~J., Lundqvist,
  P., Smith, L.~J., Pettini, M., \& King, D.~L.\ 1996, \mnras, 283,
  1355

\bibitem[Filippenko(1997)]{filippenko97} Filippenko, A.~V.\ 1997,
   ARAA, 35, 309

\bibitem[Filippenko(2005)]{filippenko05} ------ 2005, in
    White Dwarfs: Cosmological and Galactic Probes, ed. E. M. Sion, 
    S. Vennes, \& H. L. Shipman (Dordrecht: Springer), 97  

\bibitem[Filippenko et al.(2001)]{filippenko01} Filippenko, A.~V., Li,
  W.~D., Treffers, R.~R., \& Modjaz, M.\ 2001, in ASP Conf. Ser. 246, Small
  Telescope Astronomy on Global Scales, ed. W. P. Chen, C. Lemme, 
  \& B. Paczy\'{n}ski  (San Francisco: ASP), 121

\bibitem[Hamuy et al.(2003)]{hamuy03} Hamuy, M., et al.\ 2003, \nat,
  424, 651

\bibitem[Howell et al.(2005)]{howell05} Howell, D.~A., et al.\ 2005,
  \apj, 634, 1190

\bibitem[Jha et al.(2007){Jha, Riess, \& Kirshner}]{jrk07} Jha, S.,
  Riess, A.~G., \& Kirshner, R.~P.\ 2007, \apj, 659, 122

\bibitem[Koribalski et al.(2004)]{koribalski04} Koribalski, B.~S., 
et al.\ 2004, \aj, 128, 16 

\bibitem[Leibundgut(2004)]{leibundgut04} Leibundgut, B.\ 2004, 
\apss, 290, 29 

\bibitem[Li et al.(2000)]{li00} Li, W.~D., et al.\ 2000, in AIP
  Conf. Ser. 522, Cosmic Explosions, ed. S. S. Holt \& W. W. Zhang
  (New York: AIP), 103

\bibitem[Li et al.(2001)]{li01} ------\ 2001, \pasp, 113, 1178

\bibitem[Li et al.(2003)]{li03} ------\ 2003, \pasp, 115, 453

\bibitem[Mannucci et al.(2005)]{mannucci05} Mannucci, F., Della 
Valle, M., Panagia, N., Cappellaro, E., Cresci, G., Maiolino, R., 
Petrosian, A., \& Turatto, M.\ 2005, \aap, 433, 807 

\bibitem[Mannucci et al.(2006){Mannucci, Della Valle, \&
    Panagia}]{mannucci06} Mannucci, F., Della Valle, M., \& Panagia,
  N.\ 2006, \mnras, 370, 773

\bibitem[Mattila et al.(2005)]{mattila05} Mattila, S., Lundqvist, P.,
  Sollerman, J., Kozma, C., Baron, E., Fransson, C., Leibundgut, B.,
  \& Nomoto, K.\ 2005, \aap, 443, 649

\bibitem[Miller \& Stone(1993)]{miller93} Miller, J. S., \& Stone,
   R. P. S. 1993, Lick Obs. Tech. Rep. 66

\bibitem[Nakano \& Itagaki(2007)]{ni07} Nakano, S., \& Itagaki,
  K.\ 2007, \iaucirc, 8817, 3

\bibitem[O'Brien et al.(2006)]{obrien06} O'Brien, T.~J., et al.\ 
2006, \nat, 442, 279 

\bibitem[Patat et al.(2007a)]{patat} Patat, F., et al. 2007a, Science,
  317, 924

\bibitem[Patat et al.(2007b)]{patat07b} Patat, F., et al. 2007b, \aap,
  in press (preprint at ArXiv e-prints, 708, 0708.3698)

\bibitem[Perlmutter et al.(1999)]{perlmutter99} Perlmutter, S., et
  al.\ 1999, \apj, 517, 565

\bibitem[Quimby et al.(2007){Quimby, H{\"o}flich, \&
    Wheeler}]{quimby07} Quimby, R., H{\"o}flich, P., \& Wheeler,
  J.~C.\ 2007, \apj, in press (preprint at ArXiv e-prints, 705,
  arXiv:0705.4467)

\bibitem[Riess et al.(1998)]{riess98} Riess, A.~G., et al.\ 
1998, \aj, 116, 1009 

\bibitem[Riess et al.(2007)]{riess07} Riess, A.~G., et al.\ 
2007, \apj, 659, 98 

\bibitem[Salgado et al.(2007)]{salgado07} Salgado, F., Hamuy, M,
  Morrell, N., \& Folatelli, G.\ 2007, CBET 865, 1

\bibitem[Scannapieco \& Bildsten(2005)]{sb05} Scannapieco, 
E., \& Bildsten, L.\ 2005, \apjl, 629, L85 

\bibitem[Schlegel et al.(1998){Schlegel, Finkbeiner, \& Davis}]{sfd98}
  Schlegel, D.~J., Finkbeiner, D.~P., \& Davis, M.\ 1998, \apj, 500,
  525

\bibitem[Stetson(1987)]{stetson87} Stetson, P.~B.\ 1987, \pasp, 99, 191

\bibitem[Sullivan et al.(2006)]{sullivan06} Sullivan, M., et al.\ 
2006, \apj, 648, 868 

\bibitem[Tull(1998)]{hrs} Tull, R.~G.\ 1998, \procspie, 3355, 387

\bibitem[Vogt et al.(1994)]{vogt94} Vogt, S.~S., et al.\ 1994,
  \procspie, 2198, 362

\bibitem[Wang et al.(2003)]{wang03} Wang, S., et al.\ 2003,
  \procspie, 4841, 1145

\bibitem[Wang et al.(2007)]{wang07} Wang, X., et al.\ 2007, submitted
  to \apj\ (preprint at ArXiv e-prints, 708, arXiv:0708.0140)

\bibitem[Welty \& Fitzpatrick(2001)]{wf01} Welty, D.~E., \& 
Fitzpatrick, E.~L.\ 2001, \apj, 551, L175 


\end{thebibliography}
\end{document}